\documentclass[preprint,nofootinbib,unsortedaddress,showpacs,showkeys]{revtex4}
\usepackage{amsfonts,amsmath,amssymb}

\input{tcilatex}

\begin{document}

\title{Photon Damping in One-Loop HTL Perturbation Theory}
\author{Abdessamad Abada$^{\mathrm{a,b}}$}
\email{a.abada@uaeu.ac.ae}
\author{Nac\'{e}ra Daira-Aifa$^{\mathrm{a,c}}$}
\email{daira_aifa@yahoo.fr}
\affiliation{$^{\mathrm{a}}$Laboratoire de Physique des Particules et Physique
Statistique, Ecole Normale Sup\'{e}rieure, BP 92 Vieux Kouba, 16050 Alger,
Algeria\\
$^{\mathrm{b}}$Physics Department, United Arab Emirates University, POB
17551 Al Ain, UAE\\
$^{\mathrm{c}}$Facult\'{e} de Physique, Universit\'{e} des Sciences et de la
Technologie Houari Boumediene, BP\ 32 El-Alia, Bab Ezzouar 16111 Alger,
Algeria}

\begin{abstract}
We determine the damping rates of slow-moving photons in next-to-leading
order hard-thermal-loop perturbation of massless QED. We find both
longitudinal and transverse rates finite, positive, and equal at zero
momentum. Various divergences, light-cone and at specific momenta, but not
infrared, appear and cancel systematically.
\end{abstract}

\date{\today }
\maketitle

Hard thermal loops (HTL) have been used recently in different contexts. For
example, to discuss CP asymmetries and high-temperature leptogenesis \cite%
{kiessig-plumacher}, to determine the three-loop HTL contributions to the
thermodynamic functions of a quark-gluon plasma \cite%
{strickland-andersen-leganger-su} and in finite-temperature QED \cite%
{su-andersen-strickland}, to calculate the electric and magnetic properties
of the quark-gluon plasma \cite{liu-luo-wang-xu}, to find the quark number
susceptibility at finite temperature and zero chemical potential \cite%
{haque-mustafa--jiang-zhu-sun-zong}. More theoretically-oriented work covers
trying to improve the HTL perturbation expansion \cite{haque-mustafa},
finding a recursion relation for\ special one-loop diagrams that translates
into an equation for their generating functional \cite{besak-bodeker},
calculating hard thermal loops for a spatial 't Hooft loop in the deconfined
phase of a gauge theory \cite{hidaka-pisarski}, studying the
high-temperature behavior of $n$-point thermal loops in static Yang-Mills
and gravitational fields \cite%
{frenkel-pereira-takahashi--brandt-frenkel-taylor}, studying the real-time
formulation of hard thermal loops \cite{caron-huot}, looking at a
HTL-resummed improved ladder Dyson-Schwinger equation and gauge invariance 
\cite{nakkagawa-yokota-yoshida}, and extending the formulation to $\mathcal{%
N }=4$ supersymmetric Yang-Mills theories \cite%
{blaizot-iancu-kraemmer-rebhan}.

Hard thermal loops have been the answer to early problems encountered in
standard loop-expansion in high-temperature QCD \cite{HTL-structure}. In a
line of works \cite{abada-et-al--HTL}, we have looked at the behavior in the
infrared of HTL-fully-dressed one-loop dispersion relations of slow-moving
longitudinal and transverse gluons, quarks, fermions in QED, and
quasiparticles in scalar QED. The present work finishes this line by looking
at the one-loop HTL-fully-dressed damping rates for slow-moving longitudinal
and transverse photons in small-coupling finite-temperature massless QED.
These are found finite, positive, and equal at zero-momentum. Some of the
issues encountered in the context of finite-temperature QED can be found in 
\cite{QED-before}.

In the imaginary-time formalism, the quasiparticle momentum is $P_{\mu
}=(p_{0},\mathbf{p})$ with $P^{2}=p_{0}^{2}+p^{2}$. The Matsubara frequency $%
p_{0}$ is equal to $2n\pi T$ ($\left( 2n+1\right) \pi T$ ) for photons
(fermions), $n$ an integer. Real-time amplitudes are obtained via the
analytic continuation $p_{0}=-i\omega _{\mathrm{ph}}+0^{+}$, with $\omega _{ 
\mathrm{ph}}$ the real energy of the quasiparticle. We assume a regime of
small coupling constant $e$ so that the temperature $T$ is the hard scale, $%
eT$ the soft scale. The poles of the HTL-dressed photon propagators $\Delta
_{\ell ,t}(-i\omega ,\mathbf{p})$ determine the dispersion laws $\omega
_{\ell ,t}(p)$ at lowest order in the coupling ($\ell $ for longitudinal, $t$
for transverse). We use the Feynman gauge. The lowest-order dispersion
relations write: 
\begin{equation}
-\omega _{\ell }^{2}+p^{2}+\delta \Pi _{\ell }(-i\omega _{\ell },p)=0;\qquad
-\omega _{t}^{2}+p^{2}+\delta \Pi _{t}(-i\omega _{t},p)=0.
\label{lowest-order-photon-dispersions}
\end{equation}
The quantities $\delta \Pi _{\ell }$ and $\delta \Pi _{t}$ are hard thermal
loops, given by: 
\begin{eqnarray}
\delta \Pi _{\ell }(-i\omega ,p) &=&3\omega _{0}^{2}\dfrac{\omega ^{3}}{%
p^{3} }\left( \frac{p^{2}}{\omega ^{2}}-1\right) \left[ \frac{p}{\omega }
-Q_{0}(\left( \dfrac{p}{\omega }\right) \right] ;  \notag \\
\delta \Pi _{t}(-i\omega ,p) &=&3\omega _{0}^{2}\dfrac{\omega ^{3}}{2p^{3}} %
\left[ \left( \dfrac{p^{2}}{\omega ^{2}}-1\right) Q_{0}\left( \dfrac{p}{
\omega }\right) +\dfrac{p}{\omega }\right] ,  \label{photonic-htl}
\end{eqnarray}
with $Q_{0}\left( x\right) =\frac{1}{2}\log \left( \dfrac{x+1}{x-1}\right) $
and $\omega _{0}=eT/3$ the frequency of the plasma. For soft momenta, the
dispersion laws, real at lowest order, can be determined analytically: 
\begin{equation}
\omega _{\ell }\left( p\right) =\omega _{0}\left( 1+\dfrac{3}{10}\bar{p}%
^{2}- \frac{3}{280}\bar{p}^{4}+\mathcal{\dots }\right) ;\quad \omega
_{t}\left( p\right) =\omega _{0}\left( 1+\dfrac{3}{5}\bar{p}^{2}-\frac{9}{35}%
\bar{p} ^{4}+\mathcal{\dots }\right) ,  \label{expansion-w_lt}
\end{equation}
with $\bar{p}=p/\omega _{0}$.

The HTL-one-loop dressed propagator $D_{\mu \nu }\left( P\right) $ is
obtained by adding to the lowest-order $\Delta _{\mu \nu }^{-1}\left(
P\right) $ the HTL-dressed one-loop-order photon self-energy $\Pi _{\mu \nu
}(P)$: 
\begin{equation}
D_{\mu \nu }^{-1}\left( P\right) =\,\Delta _{\mu \nu }^{-1}\left( P\right)
+\,\Pi _{\mu \nu }\left( P\right) .  \label{inverse-full-propagator}
\end{equation}
Using the transversality identity $P_{\mu }\Pi ^{\mu \nu }=0$, we can
decompose the dressed self-energy into longitudinal and transverse
components, namely $\Pi _{\mu \nu }(P)=\,\Pi _{\ell }(P)P_{\mu \nu }^{\ell
}+\,\Pi _{t}(P)P_{\mu \nu }^{t}$, with $P_{\mu \nu }^{\ell ,t}$ the
corresponding projectors, and: 
\begin{equation}
\Pi _{\ell }\left( P\right) =\dfrac{P^{2}}{p^{2}}\,\Pi _{44}\left( P\right)
;\qquad \Pi _{t}\left( P\right) =\dfrac{1}{2}\left( \Pi _{ii}\left( P\right)
-\hat{p}_{i}\hat{p}_{j}\Pi _{ij}\left( P\right) \right) .  \label{Pi-l-t}
\end{equation}
Sum over repeated indices is assumed and $\mathbf{\hat{p}}=\mathbf{p}/p$.
The damping rates $\gamma _{\ell ,t}\left( p\right) $ are the negatives of
the imaginary parts of the corresponding complex energies, poles of $D_{\ell
,t}\left( P\right) $. In HTL-dressed perturbation, they are obtained via the
relation: 
\begin{equation}
\gamma _{\ell ,t}\left( p\right) =\left. \dfrac{\text{\textrm{Im}}\Pi _{\ell
,t}\left( -i\omega ,p\right) }{-2\omega +\partial \,\delta \Pi _{\ell
,t}\left( -i\omega ,p\right) /\partial \omega }\right\vert _{\omega =\omega
_{\ell ,t}\left( p\right) }.  \label{gamma-l-t--def}
\end{equation}
The denominators in this expression are readily calculable, and we have: 
\begin{equation}
\gamma _{l}\left( p\right) =\dfrac{\omega _{0}}{2p^{2}}\left[ 1+\mathcal{O}
\left( \bar{p}^{2}\right) \right] \text{\textrm{Im}}\Pi _{44};\quad \gamma
_{t}\left( p\right) =-\dfrac{1}{4\omega _{0}}\left[ 1-\dfrac{4}{5}\bar{p}
^{2} +\mathcal{O}\left( \bar{p}^{4}\right) \right] \text{\textrm{Im}}\left(
\Pi _{ii}-\hat{p}_{i}\hat{p}_{j}\Pi _{ij}\right) .  \label{gamma-l-t--def2}
\end{equation}

The HTL-dressed self energy $\Pi $ is the sum of two one-loop diagrams $\Pi
^{(1)}$ and $\Pi ^{(2)}$ given by the relations: 
\begin{align}
\Pi _{\mu \nu }^{\left( 1\right) }\left( P\right) & =e^{2}\mathrm{Tr}_{ 
\mathrm{soft}}\mathrm{Tr}_{\gamma }\left[ \Gamma _{\mu }(K-P,\,-K;\,P)\Delta
_{f}(K)\Gamma _{\nu }(K,\,-K+P;\,-P)\Delta _{f}(K-P)\right] ;  \notag \\
\Pi _{\mu \nu }^{\left( 2\right) }\left( P\right) & =e^{2}\mathrm{Tr}_{ 
\mathrm{soft}}\mathrm{Tr}_{\gamma }\left[ \Gamma _{\mu \nu
}(K,\,-K;\,P,\,-P)\Delta _{f}(K)\right] .  \label{Pi1-Pi2--def}
\end{align}
In these relations, $\mathrm{Tr}_{\mathrm{soft}}\equiv T\underset{n}{\sum }
\int d^{3}k/\left( 2\pi \right) ^{3}$ and `soft' means only soft values of
the loop momentum $k$ are to be included. Also, $\mathrm{Tr}_{\gamma }$\
means tracing over Dirac gamma matrices. The amplitude $\Pi ^{(1)}$ is the
equivalent in HTL perturbation of the classic fermionic loop whereas $\Pi
^{\left( 2\right) }$ is a pure HTL\ effect. The propagator in (\ref%
{Pi1-Pi2--def}) is fermionic, writing as: 
\begin{equation}
\Delta _{f}^{-1}\left( P\right) =\Delta _{0f}^{-1}\left( P\right) -\delta
\Sigma _{f}\left( P\right) ;\quad \delta \Sigma _{f}\left( P\right)
=m_{f}^{2}\int \frac{d\Omega _{s}}{4\pi }\frac{S\hspace{-0.6137pc}/}{PS}.
\label{delta_f}
\end{equation}
The quantity $\Delta _{0f}$ is the undressed fermionic propagator and $%
\delta \Sigma $ a hard thermal loop. The quantity $m_{f}=\sqrt{1/8}\,eT$ is
the thermal fermion mass and $S=\left( i,\mathbf{\hat{s}}\right) $. We can
decompose $\Delta _{f}$ along helicity eigenstates: 
\begin{equation}
\Delta _{f}\left( P\right) =-\left[ \mathbf{\gamma }_{+p}\Delta _{+}\left(
P\right) +\mathbf{\gamma }_{-p}\Delta _{-}\left( P\right) \right] .
\label{delta-f2}
\end{equation}
Here, $\mathbf{\gamma }_{\pm p}=\left( \gamma _{0}\pm i\mathbf{\gamma }. 
\mathbf{\hat{p}}\right) /2$ are the two helicity-state projectors and $%
\gamma _{\mu }$ the euclidean Dirac matrices. The two propagators $\Delta
_{\pm }$ are equal to $\left( D_{0}\mp D_{s}\right) ^{-1}$ respectively,
with: 
\begin{equation}
D_{0}\left( P\right) =ip_{0}-\frac{m_{f}^{2}}{p}Q_{0}\left( \frac{ip_{0}}{p}
\right) ;\quad D_{s}\left( P\right) =p+\frac{m_{f}^{2}}{p}\left[ 1-\frac{
ip_{0}}{p}Q_{0}\left( \frac{ip_{0}}{p}\right) \right] .  \label{D0-Ds}
\end{equation}

\noindent The HTL-dressed vertices $\Gamma $ involved in (\ref{Pi1-Pi2--def}
) are as follows: 
\begin{eqnarray}
\Gamma _{\mu }(K-P,\,-K;\,P) &=&\gamma _{\mu }+m_{f}^{2}\int \frac{d\Omega
_{s}}{4\pi }\frac{S_{\mu }S\hspace{-0.6137pc}/}{PS\,QS}\,;  \notag \\
\Gamma _{\mu \nu }(K,-K;P,-P) &=&-2m_{f}^{2}\int \frac{d\Omega _{s}}{4\pi } 
\frac{S_{\mu }S_{\nu }S\hspace{-0.616pc}/}{PS\,\left( P+K\right) S\,\left(
P-K\right) S}.  \label{dressed-vertices}
\end{eqnarray}

In view of (\ref{gamma-l-t--def2}), we only need the quantities $\Pi _{44}$, 
$\Pi _{ii}$ and $\hat{p}_{i}\hat{p}_{j}\Pi _{ij}$. We write these for the
first amplitude in (\ref{Pi1-Pi2--def}) as follows: 
\begin{align}
\Pi _{44}^{\left( 1\right) }& =-\frac{e^{2}}{4}\mathrm{Tr}_{\mathrm{soft}} 
\underset{\varepsilon ,\eta }{\sum }\left[ 1+\varepsilon \eta \mathbf{\hat{k}
\hat{q}}+2I_{0}+2\varepsilon I_{3}+2\eta I_{q}+2\varepsilon \eta \mathbf{\ 
\hat{k}\hat{q}}I_{0}\right.  \notag \\
& +\left. 2\varepsilon I_{0}I_{3}+2\eta I_{0}I_{q}+2\varepsilon \eta
I_{3}I_{q}+I_{0}^{2}+I_{2}^{2}+I_{3}^{2}+\varepsilon \eta \mathbf{\hat{k} 
\hat{q}}\left( I_{0}^{2}-I_{2}^{2}-I_{3}^{2}\right) \right] \Delta
_{\varepsilon }\Delta _{\eta }^{^{\prime }};  \label{Pi1_44}
\end{align}
\begin{align}
\Pi _{ii}^{\left( 1\right) }& =\frac{e^{2}}{4}\mathrm{Tr}_{\mathrm{soft}} 
\underset{\varepsilon ,\eta }{\sum }\left[ 3-\varepsilon \eta \,\mathbf{\hat{
k}\hat{q}}-2\varepsilon I_{3}-2I_{0}+2\varepsilon \eta \mathbf{\hat{k}\hat{q}
}I_{0}-2\eta I_{q}-4\varepsilon \eta I_{3q}\right.  \notag \\
& +2\varepsilon I_{2}I_{23}+2\varepsilon
I_{3}I_{33}+I_{2}^{2}+I_{3}^{2}+I_{11}^{2}+I_{22}^{2}+2I_{23}^{2}+I_{33}^{2}+2\eta I_{2}I_{2q}+2I_{3}I_{3q}
\notag \\
& +\left. 2\varepsilon \eta I_{23}I_{2q}+2\varepsilon \eta
I_{33}I_{3q}-\varepsilon \eta \mathbf{\hat{k}\hat{q}}\left(
I_{11}^{2}+I_{22}^{2}-I_{2}^{2}-I_{3}^{2}+2I_{23}^{2}+I_{33}^{2}\right) %
\right] \Delta _{\varepsilon }\Delta _{\eta }^{^{\prime }};  \label{Pi1_ii}
\end{align}
\begin{align}
\hat{p}_{i}\hat{p}_{j}\Pi _{ij}^{\left( 1\right) }& =\frac{e^{2}}{4}\mathrm{%
\ \ Tr}_{\mathrm{soft}}\underset{\varepsilon ,\eta }{\sum }\left[
1-\varepsilon \eta \mathbf{\hat{k}\hat{q}}+2\varepsilon \eta x\mathbf{\hat{q}
\hat{p}} \right.  \notag \\
& -2\eta \mathbf{\hat{q}\hat{p}}I_{p}-2\varepsilon
xI_{p}-2I_{pp}-2\varepsilon \eta \mathbf{\hat{q}\hat{p}}\,I_{p3}+2
\varepsilon \eta \mathbf{\hat{k}\hat{q}}\,I_{pp}-2\varepsilon \eta xI_{qp} 
\notag \\
& +\left. I_{p}^{2}+2\varepsilon I_{p}I_{p3}+2\eta I_{p}I_{qp}+2\varepsilon
\eta I_{p3}I_{qp}+I_{pi}^{2}-\varepsilon \eta \mathbf{\hat{k}\hat{q}}\left(
I_{pi}^{2}-I_{p}^{2}\right) \right] \Delta _{\varepsilon }\Delta _{\eta
}^{^{\prime }}.  \label{Pi1_ij}
\end{align}
In these expressions, $\Delta _{\varepsilon }\equiv \Delta _{\varepsilon
}\left( K\right) $ and $\Delta _{\eta }^{\prime }\equiv \Delta _{\eta
}\left( Q\right) $ with $Q=K-P$. Also, the summation indices $\left(
\varepsilon ,\eta \right) $ belong to the set $\left\{ \left( +,+\right)
,\left( -,-\right) ,\left( +,-\right) ,\left( 0,+\right) ,\left( 0,-\right)
\right\} $ with $\Delta _{0}\left( K\right) =Q_{0}\left( ik_{0}/k\right) $.
Furthermore, $x=\mathbf{\hat{p}\hat{k}}$. The quantities $I_{\alpha \beta
\dots }$ are solid-angle integrals given in (\ref{solid-angle-I-J}) below.
For the second amplitude in (\ref{Pi1-Pi2--def}), we have similar
expressions: 
\begin{eqnarray}
\Pi _{44}^{\left( 2\right) }\left( P\right) &=&-\,\Pi _{ii}^{\left( 2\right)
}\left( P\right) =-e^{2}\mathrm{Tr}_{\mathrm{soft}}\underset{\varepsilon
=\pm }{\sum }\left[ \left( \Delta _{\varepsilon }^{^{\prime }}-\Delta
_{\varepsilon }\right) J_{o}+\varepsilon \Delta _{\varepsilon }^{^{\prime
}}J_{q}-\varepsilon \Delta _{\varepsilon }J_{3}\right] ;  \notag \\
\hat{p}_{i}\hat{p}_{j}\Pi _{ij}^{\left( 2\right) }\left( P\right) &=&e^{2} 
\mathrm{Tr}_{\mathrm{soft}}\underset{\varepsilon }{\sum }\left[ \left(
\Delta _{\varepsilon }^{\prime }-\Delta _{\varepsilon }\right)
J_{pp}+\varepsilon \Delta _{\varepsilon }^{^{\prime }}J_{ppq}-\varepsilon
\Delta _{\varepsilon }J_{pp3}\right] .  \label{Pi2_all}
\end{eqnarray}
The quantities $J_{\alpha \beta \dots }$ stand also for solid-angle
integrals, defined here: 
\begin{equation}
I_{\alpha \beta \dots }=m_{f}^{2}\dint \dfrac{d\Omega _{s}}{4\pi }\dfrac{ 
\mathbf{\hat{\alpha}\mathbf{\hat{s}}\;\hat{\beta}\hat{s}}\dots }{QS\,KS}
;\qquad J_{\alpha \beta \dots }=m_{f}^{2}\dint \dfrac{d\Omega _{s}}{4\pi } 
\dfrac{\mathbf{\hat{\alpha}\mathbf{\hat{s}}\;\hat{\beta}\hat{s}}\dots }{
PS^{2}\,KS}.  \label{solid-angle-I-J}
\end{equation}
For both integrals, the indices $\alpha ,\beta ,\dots $ correspond to the
directions $\mathbf{\hat{x}},\mathbf{\hat{y}},\mathbf{\hat{z}}, \mathbf{\hat{%
p}},\mathbf{\hat{k}}$ and $\mathbf{\hat{q}}$. The subscript 0 means `one' in
the numerator of both integrands.

To proceed, it is best to work out with some detail one typical term. Let us
therefore consider the following contribution to longitudinal $\Pi
_{44}^{\left( 1\right) }$, namely the $\{+-\}$ term in the sum (\ref{Pi1_44}
). We write: 
\begin{align}
\Pi ^{+\,-}(P)& \equiv \Pi _{S}(P)+\Pi _{C}(P);  \notag \\
\Pi _{S}(P)& =-\frac{e^{2}}{4}\mathrm{Tr}_{\mathrm{soft}}\left[ 1-z+2\left(
1-z\right) I_{0}+2I_{3}-2I_{q}\right] \,\Delta _{+}\Delta _{-}^{^{\prime }};
\notag \\
\Pi _{C}(P)& =-\frac{e^{2}}{4}\mathrm{Tr}_{\mathrm{soft}}\left[
2I_{0}I_{3}-2I_{0}I_{q}-2I_{3}I_{q}+\left( 1-z\right) I_{0}^{2}+\left(
1+z\right) \left( I_{1}^{2}+I_{3}^{2}\right) \right] \,\Delta _{+}\Delta
_{-}^{^{^{\prime }}}\,.  \label{Pi-plus-minus}
\end{align}
In these expressions, we have replaced $\mathbf{\hat{k}\hat{q}}$ by $%
z=\left( k-px\right) /q$. Also, from now on, we take $\omega _{0}\equiv 1$
to ease intermediary notation. It is easy to show that the integral $I$ in ( %
\ref{solid-angle-I-J}) is the difference $b-a$ where: 
\begin{equation}
a_{\alpha \beta \dots }=m_{f}^{2}\dint \dfrac{d\Omega _{s}}{4\pi }\dfrac{ 
\mathbf{\hat{\alpha}\mathbf{\hat{s}}\;\hat{\beta}\hat{s}}\dots }{PS\,KS}
;\qquad b_{\alpha \beta \dots }=m_{f}^{2}\dint \dfrac{d\Omega _{s}}{4\pi } 
\dfrac{\mathbf{\hat{\alpha}\mathbf{\hat{s}}\;\hat{\beta}\hat{s}}\dots }{
PS\,QS},  \label{a-b}
\end{equation}
so that, with proper replacements, we get the result: 
\begin{equation}
\Pi _{S}(P)=-\frac{e^{2}}{4}\mathrm{Tr}_{\mathrm{soft}}\left[ 1-z-4\left(
1-z\right) a_{0}-4a_{3}+4a_{q}\right] \,\Delta _{+}\Delta _{-}^{^{^{\prime
}}},  \label{Pi_S}
\end{equation}
and the results: 
\begin{eqnarray}
\Pi _{C}(P) &=&\Pi _{C}^{aa}(P)+\Pi _{C}^{bb}(P)+\Pi _{C}^{ab}(P);  \notag \\
\Pi _{C}^{aa}(P) &=&-\frac{e^{2}}{4}\mathrm{Tr}_{\mathrm{soft}}\left[
2a_{0}a_{3}-2a_{0}a_{q}-2a_{3}a_{q}+\left( 1-z\right) a_{0}^{2}+\left(
1+z\right) \left( a_{1}^{2}+a_{3}^{2}\right) \right] \,\Delta _{+}\Delta
_{-}^{^{^{\prime }}};  \notag \\
\Pi _{C}^{bb}(P) &=&-\frac{e^{2}}{4}\mathrm{Tr}_{\mathrm{soft}}\left[
2b_{0}b_{3}-2b_{0}b_{q}-2b_{3}b_{q}+\left( 1-z\right) b_{0}^{2}+\left(
1+z\right) \left( b_{1}^{2}+b_{3}^{2}\right) \right] \,\Delta _{+}\Delta
_{-}^{^{^{\prime }}};  \notag \\
\Pi _{C}^{ab}(P) &=&\frac{e^{2}}{2}\mathrm{Tr}_{\mathrm{soft}}\left[
a_{0}b_{3}+a_{3}b_{0}-a_{0}b_{q}-a_{q}b_{0}-a_{3}b_{q}-a_{q}b_{3}\right. 
\notag \\
&&\;\left. +a_{0}b_{0}\left( 1-z\right) +\left( a_{1}b_{1}+a_{3}b_{3}\right)
\left( 1+z\right) \right] \,\Delta _{+}\Delta _{-}^{^{^{\prime }}}\,.
\label{Pi_C}
\end{eqnarray}
The angular integrals involved in $a_{\alpha }$ can be obtained in a power
series in $p$. In the longitudinal case and up to order $p^{2}$, we have the
following results: 
\begin{align}
a_{0}& =\dfrac{9}{8k}\left[ -xp+\frac{r}{2}\left( 1-3x^{2}\right)
p^{2}+\left( 1+rxp+\frac{1}{10}\left( 15r^{2}x^{2}-5x^{2}-5r^{2}+2\right)
p^{2}\right) Q_{0}+\dots \right] ;  \notag \\
a_{1}& =\dfrac{9}{8k}\left[ -\frac{r}{2}p-\left( 3r^{2}-\frac{2}{3}\right)
xp^{2}+\left( \frac{1}{2}p+rxp^{2}\right) \left( r^{2}-1\right) Q_{0}+\dots %
\right] \sqrt{1-x^{2}};  \notag \\
a_{2}& =0;  \notag \\
a_{3}& =\frac{9}{8k}\left[ 1+rxp+\frac{1}{30}\left(
45r^{2}x^{2}-15r^{2}+1\right) p^{2}-[r+r^{2}xp\right.  \notag \\
& +\left. \frac{r}{10}\left( 15r^{2}x^{2}-5x^{2}-5r^{2}+2\right)
p^{2}]Q_{0}+\dots \right] ;  \notag \\
a_{p}& =\frac{9}{8k}x-\frac{9}{16k}r\left( 1-3x^{2}\right) p+\frac{3}{80k}
\left( 15r^{2}-4\right) \left( 5x^{2}-3\right) xp^{2}  \notag \\
& -\frac{9}{8k}\left[ rx+\frac{1}{2}\left( 3r^{2}x^{2}-x^{2}-r^{2}+1\right)
p+\frac{1}{10}\left( 5r^{2}-3\right) r\left( 5x^{2}-3\right) xp^{2}\right]
Q_{0}+\dots ;  \notag \\
a_{q}& =\dfrac{k}{q}a_{3}-\dfrac{p}{q}a_{p}.  \label{a-alpha}
\end{align}
In these expressions, $r=ik_{0}/k$ and $Q_{0}=Q_{0}\left( r\right) $. The
angular integrals $b_{\alpha }$ in (\ref{a-b}) can be obtained from these
with suitable replacements and further expansion. Plugging these expressions
in (\ref{Pi_S}), we obtain\footnote{%
In these longitudinal calculations, the zeroth-order contributions cancel as
they should and hence are not shown. Only $\mathcal{O}\left( p^{2}\right) $
contributions are kept and displayed. In view of (\ref{gamma-l-t--def2}),
this will yield $\gamma _{\ell }(0).$}: 
\begin{equation}
\Pi _{S}\left( P\right) =e^{2}p^{2}\mathrm{Tr}_{\mathrm{soft}}\frac{\left(
1-x^{2}\right) }{16k^{3}}\left[ \left( 9-9kr-2k\right) -\left(
8k^{2}r-8k^{2}-9\right) \left( k+kr-1\right) \right] \Delta _{+}\Delta
_{-}^{^{^{\prime }}}.  \label{Pi_S-2}
\end{equation}
In this expression, terms involving the product of three functions that will
necessitate later a spectral decomposition, heavy on the extraction of the
imaginary part, have been worked out using the following identity: 
\begin{equation}
Q_{0}\Delta _{+}=\left( \dfrac{1}{r-1}-\dfrac{8k^{2}}{9}\right) \Delta _{+}+ 
\dfrac{8k}{9\left( r-1\right) }.  \label{identity-Q0-Delta}
\end{equation}
Similar work is done for $\Pi _{C}$. The $aa$ and $bb$ contributions, making
suitable changes in the latter, yield together the result: 
\begin{eqnarray}
\Pi _{C}^{aa}\left( P\right) +\Pi _{C}^{bb}\left( P\right) &=&e^{2}p^{2} 
\mathrm{Tr}_{\mathrm{soft}}\frac{\left( 1-x^{2}\right) }{256k^{4}}\left[
\left( -81\left( kr-1\right) ^{2}+18\left( 8k^{2}-8k^{2}r+9\right) \right.
\right.  \notag \\
&&\hspace{-0.5in}\left. \times \left( 1-kr-k\right) \left( 1-kr\right)
-\left( 8k^{2}r-8k^{2}-9\right) ^{2}\left( k+kr-1\right) ^{2}\right) \Delta
_{+}\Delta _{-}^{^{^{\prime }}}  \notag \\
&&.\left. +72k\left( k+kr-1\right) ^{2}\left( r-1\right) Q_{0}\Delta
_{-}^{^{^{\prime }}}\right] .  \label{Pi_C-aa-bb}
\end{eqnarray}
However, the third contribution $\Pi _{C}^{ab}$ requires more work now that $%
a$ and $b$ integrals are side by side. Still, this can be done and we find: 
\begin{equation}
\Pi _{C}^{ab}=e^{2}p^{2}\mathrm{Tr}_{\mathrm{soft}}\frac{\left(
1-x^{2}\right) }{256k^{2}}\left( 8k-8kr-8k^{2}-9+8k^{2}r^{2}\right)
^{2}\Delta _{+}\Delta _{-}^{^{^{\prime }}}\,.  \label{Pi_C-ab}
\end{equation}
Putting all these contributions together as dictated in (\ref{Pi-plus-minus}
), we end up with: 
\begin{align}
& \Pi ^{+\,-}\left( P\right) =-e^{2}p^{2}\mathrm{Tr}_{\mathrm{soft}}\left(
1-x^{2}\right) \left[ \frac{1}{128k^{2}}\left(
8k-8kr-8k^{2}-5+8k^{2}r^{2}\right) ^{2}\Delta _{+}\Delta _{-}^{^{^{\prime
}}}\right.  \notag \\
& \qquad \qquad \qquad \qquad \left. -\frac{9}{32k^{3}}\left( k+kr-1\right)
^{2}\left( r-1\right) Q_{0}\Delta _{-}^{^{^{\prime }}}\right] .
\label{Pi-plus-minus-2}
\end{align}
All other terms in $\Pi _{44}^{(1)}$ and $\Pi _{44}^{(2)}$ are treated along
similar lines and we obtain for the longitudinal HTL-dressed one-loop
self-energy the result: 
\begin{align}
\Pi _{44}\left( P\right) & =-e^{2}p^{2}\mathrm{Tr}_{\mathrm{soft}}\left(
1-x^{2}\right) \left[ \frac{9}{32k}\left( r-1\right) ^{2}\left( 1+r\right)
Q_{0}\Delta _{+}^{^{^{\prime }}}\right. +\frac{9}{32k}\left( r+1\right)
^{2}\left( 1-r\right) Q_{0}\Delta _{-}^{^{^{\prime }}}  \notag \\
& +\frac{1}{2k^{2}}\left( 2kr-2k-1\right) ^{2}\Delta _{+}\Delta
_{+}^{^{^{\prime }}}+\frac{1}{128k^{2}}\left(
8k-8kr-8k^{2}-5+8k^{2}r^{2}\right) ^{2}\Delta _{+}\Delta _{-}^{^{^{\prime }}}
\notag \\
& +\left. \frac{1}{128k^{2}}\left( 8k^{2}r^{2}-8kr-8k^{2}-5-8k\right)
^{2}\Delta _{-}\Delta _{+}^{^{^{\prime }}}\right] .  \label{Pi_44-final}
\end{align}

The imaginary part of the above expression can safely be evaluated using the
Matsubara relation: 
\begin{align}
& \mathrm{Im}T\dsum\limits_{k_{0}}\Delta _{\varepsilon }\left(
ik_{0},k\right) \Delta _{\eta }\left( iq_{0},k\right)  \notag \\
& =-\pi \int_{-\infty }^{+\infty }d\omega \int_{-\infty }^{+\infty }d\omega
^{\prime }\left[ 1-n_{f}\left( \omega \right) -n_{f}\left( \omega ^{\prime
}\right) \right] \rho _{\varepsilon }\left( \omega ,k\right) \rho _{\eta
}\left( -\omega ^{\prime },k\right) \delta \left( 1-\omega -\omega ^{\prime
}\right) .  \label{Matsubara-formula}
\end{align}
The quantity $\rho _{\varepsilon }\left( \omega ,k\right) $ is a spectral
density and $n_{f}\left( \omega \right) $ the Fermi-Dirac distribution.
Using this relation in (\ref{Pi_44-final}) and calling on (\ref%
{gamma-l-t--def2}) yield the result: 
\begin{align}
\gamma _{\ell }\left( 0\right) & =\dfrac{e^{4}T}{64\pi }\dint_{0}^{\infty
}dk\int_{-\infty }^{+\infty }d\omega \int_{-\infty }^{+\infty }d\omega
^{\prime }\mathrm{\delta }\left( 1-\omega -\omega ^{\prime }\right)  \notag
\\
& \left[ \underset{\varepsilon =\pm }{\dsum }\left[ \frac{1}{48}\left(
\varepsilon 8k-8\omega -8k^{2}+8\omega ^{2}-5\right) ^{2}\rho _{\varepsilon
}\rho _{\varepsilon }^{\prime }\right. \right.  \notag \\
& \left. \left. +\frac{3}{4}k^{-2}\left( k+\varepsilon \omega \right)
^{2}\left( k-\varepsilon \omega \right) \rho _{0}\rho _{\varepsilon
}^{\prime }\right] +\frac{1}{3}\left( 2\omega -2k-1\right) ^{2}\rho _{+}\rho
_{-}^{\prime }\right] .  \label{gamma_l--final-analy}
\end{align}
We have set $\rho _{\eta }=\rho _{\eta }\left( \omega ,k\right) $ where $%
\eta =+,-$ or 0, and $\rho _{\eta }^{\prime }=\rho _{\eta }\left( \omega
^{\prime },k\right) $. Note that the integration over the solid-angle $%
\mathbf{\hat{k}}$ is performed.

A similar expression for the transverse damping rate $\gamma _{t}\left(
p\right) $ is longer and more tedious to obtain. We spare the reader the
details and give the result: 
\begin{equation}
\gamma _{t}\left( p\right) =\dfrac{e^{4}T}{64\pi }\dint_{0}^{\infty
}dk\int_{-\infty }^{+\infty }d\omega \int_{-\infty }^{+\infty }d\omega
^{\prime }\left[ A\left( \omega ,\omega ^{\prime },k\right) +C\left( \omega
,\omega ^{\prime },k\right) \,p^{2}+\dots \right] .
\label{gamma_t--final-analy}
\end{equation}
The coefficient $A\left( \omega ,\omega ^{\prime },k\right) $ is as follows: 
\begin{eqnarray}
A &=&\underset{\varepsilon =\pm }{\dsum }\left[ \dfrac{3}{4k^{2}}\left(
k+\varepsilon \omega \right) ^{2}\left( k-\varepsilon \omega \right) \rho
_{0}+\dfrac{1}{48}\left( -8\omega +\varepsilon 8k-8k^{2}+8\omega
^{2}-5\right) ^{2}\rho _{\varepsilon }\right] \rho _{\varepsilon }^{^{\prime
}}\delta  \notag \\
&&+\dfrac{4}{3}\left( 2\omega -2k-1\right) ^{2}\rho _{+}\rho _{-}^{^{\prime
}}\delta ,  \label{A-in-gamma_t}
\end{eqnarray}
where $\delta $ stands for $\delta \left( 1-\omega -\omega ^{\prime }\right) 
$. The coefficient $C\left( \omega ,\omega ^{\prime },k\right) $ is: 
\begin{align}
C& =\underset{\varepsilon =\pm }{\dsum }\left( C_{10\varepsilon }\rho
_{0}+C_{1\varepsilon \varepsilon }\rho _{\varepsilon }\right) \rho
_{\varepsilon }^{^{\prime }}\delta +C_{1+\,-}\rho _{+}\rho _{-}^{^{\prime
}}\delta  \notag \\
& +\underset{\varepsilon =\pm }{\dsum }\left( C_{20\varepsilon }\rho
_{0}+C_{2\varepsilon \varepsilon }\rho _{\varepsilon }\right) \partial
_{k}\rho _{\varepsilon }^{^{\prime }}\delta +C_{2+\,-}\rho _{+}\partial
_{k}\rho _{-}^{^{\prime }}\delta  \notag \\
& +\underset{\varepsilon =\pm }{\dsum }\left( C_{30\varepsilon }\rho
_{0}+C_{3\varepsilon \varepsilon }\rho _{\varepsilon }\right) \partial
_{k}^{2}\rho _{\varepsilon }^{^{\prime }}\delta +C_{3+\,-}\rho _{+}\partial
_{k}^{2}\rho _{-}^{^{\prime }}\delta  \notag \\
& +\underset{\varepsilon =\pm }{\dsum }\left( C_{40\varepsilon }\rho
_{0}+C_{4\varepsilon \varepsilon }\rho _{\varepsilon }\right) \rho
_{\varepsilon }^{^{\prime }}\partial _{\omega }\delta +C_{4+\,-}\rho
_{+}\rho _{-}^{^{\prime }}\partial _{\omega }\delta .  \label{C-in-gamma_t}
\end{align}
In this relation, $\partial _{k(\omega )}$ stands for partial $\partial
/\partial k(\omega )$. The different coefficients in $C$ above are as
follows. Those for terms without derivatives are: 
\begin{align}
C_{10\varepsilon }& =-\frac{3\varepsilon \left( \omega ^{2}-k^{2}\right)
^{2} }{40k^{4}}\left( 6\omega ^{2}-6k^{2}-7\varepsilon k^{3}-3\omega
+3\omega ^{3}+3\varepsilon k\omega ^{2}-7k^{2}\omega \right) ;  \notag \\
C_{1\varepsilon \varepsilon }& =\tfrac{1}{960k^{2}}\left( 406\varepsilon
k-758\omega +1248\varepsilon k\omega -1412k^{2}+1760\varepsilon
k^{3}-1408k^{4}\right.  \notag \\
& +1408\varepsilon k^{5}-1280k^{6}+156\omega ^{2}+1312\omega ^{3}+512\omega
^{4}-1920\omega ^{5}+768\omega ^{6}-864\varepsilon k\omega ^{2}  \notag \\
& -2208k^{2}\omega -768\varepsilon k\omega ^{3}+768\varepsilon k^{3}\omega
+384\varepsilon k\omega ^{4}-2944k^{4}\omega +896k^{2}\omega
^{2}+4864k^{2}\omega ^{3}  \notag \\
& \left. -1792\varepsilon k^{3}\omega ^{2}-2816k^{2}\omega
^{4}+3328k^{4}\omega ^{2}+143\right) ;  \notag \\
C_{1+\,-}& =\tfrac{1}{240k^{2}}\left( 224\omega -112k-800k\omega
+720k^{2}+256k^{3}-512k^{4}+720\omega ^{2}-1792\omega ^{3}\right.  \notag \\
& \left. +1024\omega ^{4}+3072k\omega ^{2}-1536k^{2}\omega -2048k\omega
^{3}+1024k^{3}\omega +512k^{2}\omega ^{2}-83\right) .  \label{C1s-in-C}
\end{align}
The coefficients involving $\partial _{k}\rho _{\varepsilon }\left( \omega
^{\prime },k\right) $ are given here: 
\begin{eqnarray}
C_{20\varepsilon } &=&-\dfrac{9}{40k^{3}}\left( 2\omega -1\right) \left(
k+\varepsilon \omega \right) ^{2}\left( \text{ }k-\varepsilon \omega \right)
;  \notag \\
C_{2\varepsilon \varepsilon } &=&-\dfrac{1}{480k}\left( 10\omega
-24\varepsilon k-16\varepsilon k\omega +88k^{2}-72\omega ^{2}+48\omega
^{3}\right.  \notag \\
&&\left. -48k^{2}\omega +7\right) \left( \varepsilon 8k-8\omega
-8k^{2}+8\omega ^{2}-5\right) ;  \notag \\
C_{2+\,-} &=&-\dfrac{1}{15k}\left( 16k-16\omega -16k\omega +16\omega
^{2}+11\right) \left( 2\omega -2k-1\right) ,  \label{C2s-in-C}
\end{eqnarray}
and those involving $\partial ^{2}\rho _{\varepsilon }\left( \omega ^{\prime
},k\right) /\partial k^{2}$ write like this: 
\begin{eqnarray}
C_{30\varepsilon } &=&-\dfrac{3}{20k^{2}}\left( \varepsilon k+\omega \right)
^{2}\left( \varepsilon \omega -k\right) ;  \notag \\
C_{3\varepsilon \varepsilon } &=&\dfrac{1}{240}\left( \varepsilon 8k-8\omega
-8k^{2}+8\omega ^{2}-5\right) ^{2};\quad C_{3+\,-}=\dfrac{2}{15}\left(
2\omega -2k-1\right) ^{2}.  \label{C3s-in-C}
\end{eqnarray}
Finally, the coefficients involving $\partial _{\omega }\delta \left(
1-\omega -\omega ^{\prime }\right) $ are as follows: 
\begin{eqnarray}
C_{40\varepsilon } &=&\dfrac{9}{20k^{2}}\varepsilon \left( \varepsilon
k+\omega \right) ^{2}\left( \varepsilon k-\omega \right) ;  \notag \\
C_{4\varepsilon \varepsilon } &=&\dfrac{1}{80}\left( \varepsilon 8k-8\omega
-8k^{2}+8\omega ^{2}-5\right) ^{2};\quad C_{4+\,-}=\dfrac{4}{5}\left(
2\omega -2k-1\right) ^{2}.  \label{C4s-in-C}
\end{eqnarray}
Remember that only soft values of $k$, $\omega $ and $\omega ^{\prime }$ are
allowed, which means we have used the approximation $n_{f}\left( \omega
\right) \simeq \frac{1}{2}\left( 1-\frac{\omega }{2T}+\dots \right) $. Also,
the expansions in (\ref{expansion-w_lt}) are used to appropriate orders for
each step of the calculations. Finally, we are still keeping $\omega _{0}=1$.

Now what remains is to use the explicit expressions of the spectral
densities and perform the integrals in (\ref{gamma_l--final-analy}) and (\ref%
{gamma_t--final-analy}). These are given in \cite{spectral-functions}: 
\begin{align}
\rho _{\pm }\left( \omega ,k\right) & =\mathfrak{z}_{\pm }(k)\delta \left(
\omega -\omega _{\pm }(k)\right) +\mathfrak{z}_{\mp }(k)\delta \left( \omega
+\omega _{\pm }\left( k\right) \right) +\beta _{\pm }\left( \omega ,k\right)
\Theta \left( k^{2}-\omega ^{2}\right) ;  \notag \\
\rho _{0}\left( \omega ,k\right) & =-\frac{1}{2}\Theta \left( k^{2}-\omega
^{2}\right) .  \label{spectral-densities}
\end{align}
The residue functions $\mathfrak{z}_{\varepsilon }(k)$ at the fermionic
poles $\omega _{\varepsilon }\left( k\right) $ and the cuts $\beta
_{\varepsilon }\left( \omega ,k\right) $ are known: 
\begin{eqnarray}
\mathfrak{z}_{\pm }\left( k\right) &=&\frac{k^{2}-\omega _{\pm }^{2}\left(
k\right) }{2m_{f}^{2}};  \notag \\
\beta _{\pm }\left( \omega ,k\right) &=&\dfrac{-m_{f}^{2}\left( k\mp \omega
\right) }{2k^{2}\left[ \left[ \omega \mp k\mp \dfrac{m_{f}^{2}}{k}-\dfrac{
m_{f}^{2}}{k}\left( 1\mp \dfrac{\omega }{k}\right) Q_{0}\left( k/\omega
\right) \right] ^{2}+\dfrac{\pi ^{2}m_{f}^{4}}{4k^{2}}\left( 1\mp \dfrac{
\omega }{k}\right) ^{2}\right] }.  \label{residue-cut}
\end{eqnarray}
As before, it is best to illustrate the coming steps with a generic example.
Take then for instance the integral: 
\begin{equation}
I_{++}=\int_{0}^{+\infty }dk\int_{-\infty }^{+\infty }d\omega \int_{-\infty
}^{+\infty }d\omega ^{^{\prime }}f\left( \omega ,k\right) \rho _{+}\left(
\omega ,k\right) \partial _{k}^{2}\rho _{+}\left( \omega ^{^{\prime
}},k\right) \delta \left( 1-\omega -\omega ^{^{\prime }}\right) .
\label{I-plus}
\end{equation}
Here the integral over $\omega ^{\prime }$ is readily done with the
replacement $\omega ^{\prime }=1-\omega $. Imposing the pole conditions on
both $\omega $ and $1-\omega $ forbids `pole-pole' terms, so that we get: 
\begin{align}
I_{++}& =I_{++}^{PC}+I_{++}^{CP}+I_{++}^{CC};  \notag \\
I_{++}^{PC}& =\int_{0}^{+\infty }dk\int_{-\infty }^{+\infty }d\omega f\left(
\omega ,k\right) \mathfrak{z}_{+}\delta (\omega -\omega _{+})\partial
_{k}^{2}\left[ \beta _{+}\left( 1-\omega ,k\right) \Theta (k^{2}-\left(
1-\omega \right) ^{2})\right] ;  \notag \\
I_{++}^{CP}& =\int_{0}^{+\infty }dk\int_{-\infty }^{+\infty }d\omega f\left(
\omega ,k\right) \beta _{+}(\omega ,k)\Theta (k^{2}-\omega ^{2})\partial
_{k}^{2}\left[ \mathfrak{z}_{+}\left( k\right) \delta (1-\omega -\omega
_{+}) \right] \mathrm{;}  \notag \\
I_{++}^{CC}& =\int_{0}^{+\infty }dk\int_{-\infty }^{+\infty }d\omega f\left(
\omega ,k\right) \beta _{+}(\omega ,k)\Theta (k^{2}-\omega ^{2})  \notag \\
& \hspace{1.5in}\times \partial _{k}^{2}\left[ \beta _{+}\left( 1-\omega
,k\right) \Theta (k^{2}-\left( 1-\omega \right) ^{2})\right] \mathrm{.}
\label{I-plus-details}
\end{align}
The usual care must be taken when manipulating the Heaviside distribution
and its derivatives. For the `pole-cut' contribution, we obtain the result: 
\begin{eqnarray}
I_{++}^{PC} &=&\dint\limits_{k_{p}}^{\infty }dk\mathfrak{z}_{+}f\left(
\omega _{+},k\right) \left. \partial _{k}^{2}\beta _{+}\left( \omega
,k\right) \right\vert _{\omega =1-\omega _{+}}+\left[ \mathfrak{z}
_{+}f\left( \omega _{+},k\right) \dfrac{\left. \partial _{k}\beta _{+}\left(
\omega ,k\right) \right\vert _{\omega =1-\omega _{+}}}{\left\vert 1-\omega
_{+}^{\prime }\right\vert }\right] _{k=k_{p}}  \notag \\
&=&-0.031503.  \label{I-plus-PC}
\end{eqnarray}
In this expression, $k_{p}$ is the solution of the equation $\omega
_{+}\left( k\right) =1+k$. Also, the prime here means derivative with
respect to $k$. The contribution `cut-pole' is given in the following
expression: 
\begin{eqnarray}
I_{++}^{CP} &=&\dint\limits_{k_{p}}^{\infty }dk\left[ f\left( \omega
,k\right) \beta _{+}\left( \omega ,k\right) \mathfrak{z}_{+}^{\prime \prime
}-\left( 2\mathfrak{z}_{+}^{\prime }\omega _{+}^{\prime }+\mathfrak{z}
_{+}\omega _{+}^{\prime \prime }\right) \partial _{\omega }\left[ f\left(
\omega ,k\right) \beta _{+}\left( \omega ,k\right) \right] \right.  \notag \\
&&\left. +\mathfrak{z}_{+}\omega _{+}^{\prime }{}^{2}\partial _{\omega }^{2} %
\left[ f\left( \omega ,k\right) \beta _{+}\left( \omega ,k\right) \right] %
\right] _{\omega =1-\omega _{+}}  \notag \\
&&\left. +\left[ \mathfrak{z}_{+}(\omega _{+}^{\prime })^{2}f\left( 1-\omega
_{+},k\right) \dfrac{\partial _{\omega }\beta _{+}\left( \omega ,k\right)
_{\omega =1-\omega _{+}}}{\left\vert 1-\omega _{+}^{\prime }\right\vert } %
\right] _{k=k_{p}}=-2.796605.\right.
\end{eqnarray}
Note here that, because of the divergent behavior of the cut functions and
some of their derivatives at $k_{p}$, some of the terms, individually, are
singular at this point. However, taking them together carefully cancels the
singularities, and the final result is finite. Details have been presented
elsewhere \cite{ABDe}. Finally, the `cut-cut' contribution writes as: 
\begin{equation}
I_{++}^{CC}=-\dint\limits_{0.5}^{\infty }dk\dint\limits_{1-k}^{k}d\omega
\partial _{k}\left[ f\left( \omega ,k\right) \beta _{+}\left( \omega
,k\right) \right] \partial _{k}\beta _{+}\left( 1-\omega ,k\right) =-0.04266.
\label{I-plus-CC}
\end{equation}
It is finite, though with singularities on the light-cone handled with
suitable change of variables and adequate algebra \cite{ABDe}. As we see,
the integral $I_{++}$ is finite, equal to $-2.870\,8$.

Similar procedures are applied to all other terms with various lengths. When
all are summed and $\omega _{0}$ restored, the final results are the
following: 
\begin{equation}
\gamma _{\ell }\left( p\right) =\dfrac{e^{4}T}{64\pi }\left[ 1.32954+ 
\mathcal{O}(\bar{p}^{2})\right] ;\quad \gamma _{t}\left( p\right) =\dfrac{
e^{4}T}{64\pi }\left[ 1.32954+3.05041\tilde{p}^{2}+\mathcal{O}(\bar{p}^{4}) %
\right] .  \label{damping-rates--final}
\end{equation}

\noindent Remember that $\bar{p}=p/\omega _{0}$. The longitudinal and
transverse one-loop HTL-dressed damping rates are finite, positive and equal
at zero momentum. The absence of infrared sensitivity is related to the
absence of photon propagators at one-loop HTL-dressed order, which remain
massless beyond the electric (soft)\ scale. Their absence is also related to
the appearance of the one-loop HTL-dressed photonic damping rates at the
higher order $e^{3}\left( eT\right) $ instead of the next-to-leading order $%
e\left( eT\right) $, like the fermionic damping rates for example \cite{ABDe}
. Indeed, for soft $\omega $ and $\omega ^{\prime }$, the thermal factor $%
\left( 1-n_{f}\left( \omega \right) -n_{f}\left( \omega ^{\prime }\right)
\right) $ is of $\mathcal{O}\left( e\right) $ instead of $\mathcal{O}\left(
1/e\right) $ if we had soft photons in the loops.

It is important to stress that the results in (\ref{damping-rates--final})
come from the calculation of the imaginary part of the one-loop HTL-dressed
self-energy of a soft (almost static) photon in a thermalized QED\ plasma.
Loop momenta are soft. One then naturally asks: are there other diagrams not
incorporated in the \textit{one-loop} HTL summation scheme that would
contribute to the photon damping rates at this order or smaller? A similar
issue was raised previously in the literature, in the context of the
thermal-field-theory calculation of the production rate of non-thermalized
soft photons in a high-temperature quark-gluon plasma in thermal
equilibrium. When calculated in the strict\ one-loop HTL-dressed
prescription, the imaginary part of the photon self-energy $\func{Im}\Pi $,
to which the production rate is proportional, was found to be of the order $%
e^{2}g^{3}T^{2}$, but exihibiting a colinear divergence for real light-cone
emitted photons \cite{baier-et-al--aurenche-et-al}; $g$ is the QCD coupling
constant. This is a quite general feature of the hard thermal loops, which
become colinearly singular when some of their external soft momenta are put
on the light cone. This problem was addressed in \cite{flechsig-rebhan} and
resolved by endowing the internal hard fermions with an asymptotic thermal
mass $m_{\infty }$. Single-pole colinear singularities thus regularized
enhance a given order by a factor $\ln \left( T^{2}/m_{\infty }^{2}\right)
\sim \ln \left( 1/g\right) $ for soft $m_{\infty }\sim gT$. However, more
severely, because of double-pole colinear singularities, some two-loop
diagrams with hard internal fermions and a soft gluon insertion that are a
priori dismissed in the standard HTL-summation order-of-magnitude counting
are found to contribute and dominate over the mere one-loop HTL-dressed
result \cite{aurenche-gelis-kobes-petitgirard}. More precisely, when the
double-pole colinear singularity in the two-loop diagram is regularized with 
$m_{\infty }$, it contributes an enhancement factor $T^{2}/m_{\infty
}^{2}\sim 1/g^{2}$ instead of a simple log, which means an order $%
e^{2}gT^{2} $ for $\func{Im}\Pi $ for ligh-cone photons. This result carries
through to the case of virtual photons, relevant to lepton pair production,
but in the static limit $p\ll \omega _{\mathrm{ph}}$ where $\omega _{\mathrm{%
ph}}$ is the energy of the photon, such two-loop diagrams go to zero \cite%
{aurenche-gelis-kobes-petitgirard}.

The production rate of soft static photons has been evaluated at one-loop
HTL-dressed order in \cite{braaten-pisarski-yuan} and reexamined in \cite%
{aurenche-gelis-kobes-zaraket} at the two-loop level in the limit $\omega _{ 
\mathrm{ph}}\ll m_{g}\ll T$, where $m_{g}\sim gT$ is the gluon thermal mass.
The bremsstrahlung contribution to $\func{Im}\Pi $ is found to be of the
order $e^{2}g^{2}m_{g}^{2}T/\omega _{\mathrm{ph}}\ln \left(
T^{2}/m_{g}^{2}\right) $, of the same order as the one-loop HTL-dressed
result in the same regime. A similar conclusion is reached for the Compton
and annihilation processes. The same problem is reexamined in \cite%
{aurenche-carrington-marchal} for $\omega _{\mathrm{ph}}\gg gT$, and an
order $e^{2}g^{3}$ at two loops is also found.

The photon production rate problem has been examined in different limits and
approximation schemes \cite{pietzmann-thoma}. In that problem, there are two
independent small coupling constants, $e$ and $g$, and three independent
energy scales, $T$, $p$ and $\omega _{\mathrm{ph}}$. Our situation is
somewhat physically different, that of a thermalized QED plasma in which $e$
is the only small coupling constant, and $T$ and $p$ the only two
independent energy scales; the photons are thermally on-shell, with $\omega
_{\mathrm{ph}}=\omega _{t,l}\left( p\right) \sim eT$. Therefore, care must
be taken when drawing analogies between the photon production rate problem
and our situation. From this work and from previous experience \cite%
{abada-et-al--HTL} with the case $p\ll \omega _{\mathrm{ph}}$, we learn that
colinear issues do indeed arise in the midst of the calculations, but never
seriously enough to warrant the use of the asymptotic-mass improved HTL
perturbation \cite{flechsig-rebhan} so that they enhance the order of the
contribution. It could therfore be reasonable to expect that in our
situation, colinear divergences are safe enough not to enhance the order of
two-loop diagrams with hard internal momenta. However, it is also reasonable
to expect the same two-loop diagrams to contribute to order $e^{5}T^{2}$ as
we cannot find a convincing order-of-magnitude counting argument to rule
them out. But at the same time, it is also clear that only an explicit
investigation of such diagrams will determine their order and precise
contribution, as well as the nature of divergences they may or may not have.
This task is beyond the present framework and is left for a future project.
In light of this discussion, the results in (\ref{damping-rates--final})
must be seen as the contribution to the damping rates of soft photons with $%
p\ll \omega _{\mathrm{ph}}=\omega _{t,l}\left( p\right) $ coming from
one-loop HTL-dressed diagrams, a contribution that is finite in the infrared
and free from colinear divergences.

Finally, beyond the damping rates, one would like to study the
next-to-leading energies of the quasiparticles in HTL perturbation. For
this, one needs to look at the real parts of the HTL-dressed next-to-leading
order self-energies, something which is more complicated. There is already
the work \cite{schulz} which estimated the next-to-leading gluon mass. More
recently, the work \cite{carrington-et-al} estimated the next-to-leading
quark mass. It would be interesting to see if we can estimate the
next-to-leading dispersion laws for slow-moving quasiparticles in HTL
perturbation.

\begin{acknowledgments}
We thank A. Gherbi for his involvement in the early parts of this work.
\end{acknowledgments}

\end{document}